# Interaction Dependence Thermodynamical parameters of Harmonically Trapped Bose gas

**Ahmed S. Hassan, Shemi S. M. Soliman, and Emad H. Soliman**
*Department of Physics, Faculty of Science,*
*El Minia University, El Minia, Egypt*

In this paper the thermodynamical parameters of a condensed Boson gas are calculated from the partial derivative of the grand potential. In particular, the analytical expressions for some important parameters, such as the condensed fraction, specific heat, critical temperature, effective size, and release energy are investigated. The mean effects which can be altered the ideal Bose gas, such as finite size, highly anisotropic of the external potential and interatomic interaction effects are considered simultaneously. Some new characteristics of the trapped interacting Bose gases in a highly anisotropic trap are revealed. The calculated result for the condensed fraction is compared with the ongoing Stuttgart experiment for $^{52}Cr$ ( Griesmaier et al., Phys. Rev. Lett. 94, 160401(2005)) directly. Good agreement between both the theoretical and experimental data are obtained.

Among the elements that have been used to create Bose-Einstein condensation (BEC) is the Chromium, $^{52}Cr$. This element is a novel quantum gas with respect to its technical applications. Generally magnetic trapping works best with atoms having a strong magnetic moment (it is very large for chromium than for other elements), and laser cooling favors atoms with strong transitions in the visible or infrared region. Given these requirements, chromium atoms are especially good candidates for both BEC and degenerate quantum gases. So it is very important to study and investigate the relevant thermodynamical parameters for this element.

Several approaches have been used in calculating the thermodynamical quantities relevant to BEC phenomenon. One of the efficient methods for describing this phenomenon is the semiclassical approximation, i.e. which is the density of states (DOS) approach [1, 2, 3, 4, 5, 6]. Assuming that the system is described by the grand canonical ensemble, it's relevant thermodynamical quantities can be calculated directly from the partial derivative of the grand potential [2]. This motivates us to consider the related problem of the quantum statistics for a finite number of interacting Boson. In this article we want to give a systematic treatment for this problem.

In doing this we are going to extend a careful and beautiful study of Kirsten and Toms of an ideal non-relativistic Bose gas [2]. The authors there employed an approach where the sum over the discrete spectrum was replaced by an integral with an approximate DOS. However, previous studies showed that the resulting thermodynamical properties depend crucially on the choice and construction of the DOS [7, 8]. In our previous study, a systematic generalization for the DOS allowing for an essential extension of its region of applicability. In particular, it becomes possible to describe BEC of an interacting trapped Bose gas with finite size effect [9]. A general DOS formula for harmonically trapped gas is calculated directly from the degenerate property of the energy spectrum for the harmonic potential. The resulting thermodynamical parameters calculated by using this DOS is given in terms of the chemical potential. Using the calculated chemical potential from the semi-analytical solution of Gross-pitaevskii (G-P) equation the interaction effect is embodied in our approach and the resulting thermodynamical parameters are coincident with the quantum mechanical one for harmonic traps [10, 11, 12, 13, 14, 15].

As a direct application for the calculated results the condensed fraction is compared with the experimental measured data of Griesmaier et al. for $^{52}Cr$ [16, 17, 18, 19]. There is a good agreement between the calculated results and the measured experimental data. It was found that finite size effects, responsible for the deviations from the ideal gas model, are still visible for interacting gas. Finite size effects are strongly quenched in the presence of the interaction. The specific heat and the critical temperature for this system are discussed in detail. The calculated specific heat has a maximum at some certain temperature less than the transition temperature of the ideal gas, it is perfectly continuous and smooth at it's maximum. Our results also show that the effective size is proportional to $T^4$ for $T > T_0$, (high temperature) and it's proportional to $T$ for $T < T_0$. The release energy has the same temperature behavior like the effective size. This behavior can be used as a good indication for both BEC or the quantum degenerate gases.

The paper is organized as follows: Section two includes a simple model for the BEC in 3D harmonic trap, and the density of state approach. Section three is devoted to calculate the condensed fraction and specific heat. The dependence of the effective size and the release energy on the temperature is given in section four. Section five presents a short conclusion.

## Simple model for BEC

Quantitative information about the BEC can be obtained by considering the grand canonical ensemble for a system of N bosons confined by a 3D harmonic potential,

$$V_{ext}(r) = \frac{m}{2}\omega^2(x^2 + y^2 + z^2) \quad (1)$$

and it's corresponding quantized energy spectrum,

$$E_n = (n_x + n_y + n_z)\hbar\omega + \frac{3}{2}\hbar\omega \quad (2)$$

where $n = n_x + n_y + n_z = 0, 1, 2, 3,\ldots$. Within the grand canonical ensemble, the chemical potential $\mu$ and the temperature T are fixed by the requirement that the total number of particles, N is determined from

$$N = \sum_{n=0}^{\infty} g_n \frac{z\, e^{-\beta E_n}}{1 - z\, e^{-\beta E_n}} = N_0 + \sum_{n=0}^{\infty} \frac{g_n z\, e^{-\beta E_n}}{1 - z\, e^{-\beta E_n}} \quad (3)$$

where $\beta = (1/K_B T)$, and $g_n$ is the degeneracy of any level with energy $E_n$. The fugacity z is determined in terms of the chemical potential $\mu$ and the ground state energy of the harmonic potential, $z = e^{\beta(\mu - E_0)}$. Once $\mu$ has been determined, all thermodynamical relevant quantities can be calculated from the partial derivative of the grand potential q, which is the logarithm of the grand canonical partition function [15],

$$q = -\sum_n g_n \ln(1 - z e^{\beta E_n})$$

It is convenient to expand the logarithm to express the grand potential as a sum similar to Eq.(3)

$$q = q_0 + \sum_{n=1}^{\infty} \frac{g_n z\, e^{-\beta E_n}}{1 - z\, e^{-\beta E_n}} \quad (4)$$

For very large number of particles one can approximate the sum directly into ordinary integral weighted by an appropriate DOS $\rho(\varepsilon)$, i.e.

$$q = q_0 + \sum_{j=1}^{\infty} z^j \int_0^{\infty} \rho(E) e^{-j\beta E} dE \quad (5)$$

In our approach the DOS is determined from the degeneracy of the energy levels and requires that the condition $K_B T \gg \hbar \Omega$, with $\Omega = (\omega_x \omega_y \omega_z)^{1/3}$, to be satisfied.

*Accurate density of states for BEC*

The harmonic oscillator spectrum is entirely degenerate and discrete. The degeneracy of any level is given by,

$$g_n = (1/2)\, n^2 + (3/2)\, n + 1$$

and discrete such that any level is separated from other levels by an integer unit of $n\hbar\omega$ according to the relation $E'_n = E_n - E_0 = n\hbar\omega$. This means that the quantity $n = (E'_n / \hbar\omega)$ has a meaning of integer number. Thus the degeneracy of any energy level can be expressed in terms of $(E'_n / \hbar\omega)$ instead of the quantum number n,

$$g_n(E'_n/\hbar\omega) = \frac{1}{2}n^2 + \frac{3}{2}n + 1 = \frac{1}{2}\left(\frac{E'_n}{\hbar\omega}\right)^2 + \frac{3}{2}\left(\frac{E'_n}{\hbar\omega}\right) + \frac{3}{2}\left(\frac{E_0}{\hbar\omega}\right) \quad (6)$$

The last term in eq.(6) is chosen carefully, any other choice in terms of higher energy $E'_n$ is not acceptable. For a single particle system, the general DOS is given by the number of levels per unit energy level space, $\hbar\omega$, i. e.

$$\rho(E'_n) = \frac{g_n(E'_n/\hbar\omega)}{\hbar\omega} = \frac{1}{2}\frac{E'^2_n}{(\hbar\omega)^3} + \frac{3}{2}\frac{E'_n}{(\hbar\omega)^2} + \frac{2}{3}\frac{E_0}{(\hbar\omega)^2} \quad (7)$$

Where $E_0$ is the single particle ground state energy. For a harmonically trapped Bose gas Kirsten and Toms [8] pointed that the transition temperature for BEC is the temperature at which $\mu = E_0$. Thus, the accurate DOS for BEC can be obtained from Eq.(7) by replacing $E_0$ by $\mu$,

$$\rho(E'_n) = \frac{1}{2}\frac{E'^2_n}{(\hbar\omega)^3} + \frac{3}{2}\frac{E'_n}{(\hbar\omega)^2} + \frac{2}{3}\frac{\mu}{(\hbar\omega)^2} \quad (8)$$

here the chemical potential is defined as the single-particle ground state energy. Generalization of the result (8) to a many particles system is straightforward. The accurate DOS for a many particles system can be obtained from (7) by replacing $E_0$ by the total ground state energy per particle, let this energy $E'_n/N_0$, where $E'_0$ is the total ground state energy and $N_0$ is the number of particles in this level. The total number of particles (with energy (3/2) $\hbar\omega$ for each particle) in the ground state level can be calculated from the number of acceptable states in this level. However, for a many-particles system the energy levels become very dense, i.e. there is no energy level spacing. Let the energy of these continuum level is ε. In this case the number of the acceptable states for such level is given by $\hbar\omega/\varepsilon$. Moreover, any particle in these state will have energy

$$E_0 \equiv \frac{E_0'}{N_0} = \frac{E_0'}{\hbar\omega/\varepsilon} = \frac{\mu\varepsilon}{\hbar\omega} \qquad (9)$$

Note that the chemical potential is an intensive variable, i.e. $\mu = E'_0$ also. Gathering (7) and (9), in terms of ε one have,

$$\rho(\varepsilon) = \frac{1}{2}\frac{\varepsilon^2}{(\hbar\omega)^3} + \frac{\varepsilon}{(\hbar\omega)^2}\left\{\frac{3}{2} + \frac{2}{3}\frac{\mu}{\hbar\omega}\right\} \qquad (10)$$

The generalization of the above treatment to a potential with three different frequencies ($\omega_x$, $\omega_y$ and $\omega_z$) is straightforward. Follow-up Kirsten and Toms [8] yields,

$$\rho(\varepsilon) = \frac{1}{2}\frac{\varepsilon^2}{(\hbar\Omega)^3} + \frac{\varepsilon}{(\hbar\Omega)^2}\frac{\bar\omega}{\Omega}\left\{\frac{3}{2} + \frac{2}{3}\frac{\mu}{\hbar\Omega}\right\} \qquad (11)$$

Where $\Omega = (\omega_x\omega_y\omega_z)^{1/3}$; and $\bar\omega = \frac{1}{3}(\omega_x+\omega_y+\omega_z)$ are the geometrical average and the mean of the oscillator frequencies, respectively. Substituting Eq.(11) in (5) one has,

$$q = q_0 + \left(\frac{kT}{\hbar\Omega}\right)^3 g_4(z) + \frac{\bar\omega}{\Omega}\left(\frac{kT}{\hbar\Omega}\right)^2 g_3(z)\left\{\frac{3}{2} + \frac{2}{3}\frac{\mu}{\hbar\Omega}\right\} \qquad (12)$$

Where $g_n(z)$ is the usual Bose function. A very important problem here concerns the temperature dependence of the chemical potential. This dependence might be seen more clearly by considering the number of particles in the ground state with energy $(3/2)\hbar\omega$. Suppose that the condensate wave function is described by the Gross-Pitaevskii equation

$$-\frac{\hbar^2}{2m}\Psi(r) + V_{ext}(r)\Psi(r) + 2gn_1(r)\Psi(r) + g\Psi^3(r) = \mu\Psi(r) \qquad (13)$$

Where $g = 4\pi\hbar^2 a/m$, $a$ being the scattering length, and $n_1(r)$ is the average non-condensed particle distribution. The number of particles in the ground states is given by $N_0 \equiv N = \int\psi^2(r)\,d(r^3)$. An approximate semi-analytical solution for Eq.(13) can be obtained by treating the interaction perturbatively [10, 11]. To zero order in $gn_1(r)$ one has

$$\mu(N_0,T) = (1/2)\hbar\Omega(15a/a_r)^{2/5}\sqrt{N_0} \qquad (14)$$

with $a_r = \sqrt{(\hbar/m\Omega)}$ is a characteristic length for the harmonic trap. Eq.(14) provides a useful estimate of μ as a function of $N_0$ at any temperature T.

Another important scale for the chemical potential is it's value as a function in N. This value can be seen as follows. Since the number of particles in the ground state is given by

$$N_0 \equiv \frac{1}{z-1} = \frac{1}{e^{\beta(E_0'-\mu)}-1} \qquad (15)$$

for $\mu = E'_0$ one have $N_0 \to \infty$ (with $E'_0$ is the total ground state energy). For given T and particle number N it is clear from Eq.(15) that

$$(E'_0 - \mu) > \beta^{-1}\ln\frac{N+1}{N} \qquad (16)$$

and that ($E'_0-\mu$) can reach zero only in the zero temperature limit or in the limit $N\to\infty$. Since in the real sense most of the experiments have been carried out with finite number of atoms; the thermodynamic limit ($N\to\infty$) is never reached exactly. Consequently, we will consider that ($E'_0-\mu$) can reach zero only at T = 0. In this case the chemical potential can be simply estimated from Eq.(14) [12],

$$\mu(N,T=0) = (1/2)\hbar\Omega(15a/a_r)^{2/5}\sqrt{N} \qquad (17)$$

From Eq.(14) and (17) one has

$$\mu(N_0,T) = \mu(N,T=0)(N_0/N)^{2/5} \qquad (18)$$

In the following and when the grand potential q is used in calculating the thermodynamical parameters, it is convenient to use a dimensionless parameter η defined such as,

$$\frac{\mu(N_0,T)}{K_B T_0} = \frac{\mu(N,T=0)}{K_B T_0}(N_0/N)^{2/5} = \eta(N_0/N)^{2/5} = \eta(1-(T/T_0)^3)^{2/5} \qquad (19)$$

first introduced by Dalfovo et al.[12]. The condensed fraction for the ideal gas model $(N_0/N) = (1-(T/T_0)^3)$, with $T_0 = (\hbar\Omega/K_B)(N/\zeta(3))^{1/3}$ is used here. The parameter η is a scaling parameter gives the scaling behavior of all thermodynamical quantities due to interatomic interaction.

## Condensed fraction and the specific heat

In terms of the q-potential the total number of particles and the specific heat are given by

$$N = \beta^{-1} \frac{\partial q}{\partial \mu}\Big|_T$$

$$C(T) = K_B T^2 \frac{\partial U}{\partial T}\Big|_T \qquad (20)$$

with $U = K_B T^2 \, \partial q/\partial T|_z$ is the total energy for the system. Substituting from Eq.(12) in (20), one can calculate the condensed fraction,

$$\frac{N_0}{N} = 1 - t^3 - \frac{\bar{\omega}}{\Omega} t^2 \frac{\zeta(2)}{\zeta(3)} \left\{ \frac{3}{2} \left( \frac{\zeta(3)}{N} \right)^{\frac{1}{3}} + \frac{2}{3} \eta \left(1 - t^3\right)^{\frac{2}{5}} \right\} \qquad (21)$$

where $t = T/T_0$ is the reduced temperature. Eq.(21) accounted well for the finite size, and interatomic interaction correction effects simultaneously. In figure(1) the results calculated from Eq.(21) is compared with the experimental data of Griesmaier et.al.[16] for $^{52}$Cr. Based on the experimental setup parameters (trap frequencies $(\bar{\omega}/\Omega) = 3.31$, and number of atoms $N = 1.3 \times 10^8$) the best fit for the theoretical results with the experimental data is obtained for interaction parameter $\eta = 0.1$.

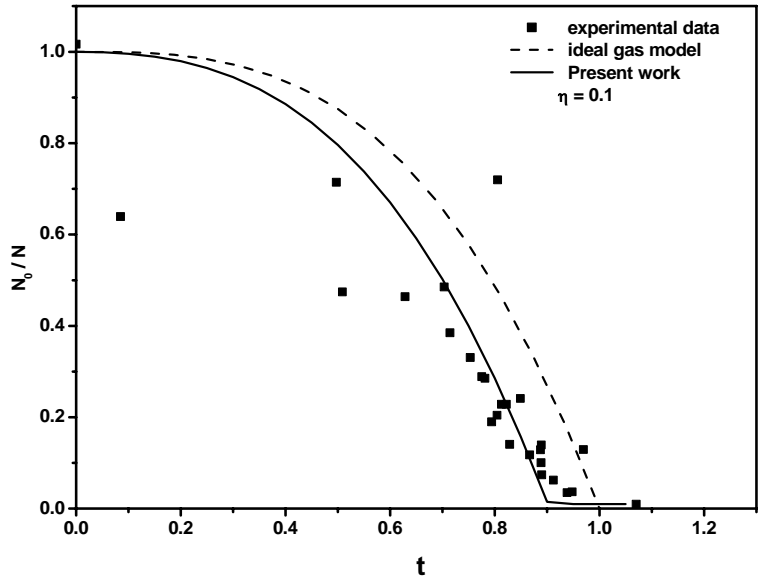

Figure 1: Condensed fraction $N_0/N$ as a function of the reduced temperature $t = T/T_0$. The results calculated from Eq.(21) for $N = 1.3 \times 10^8$.

Figure(2) is devoted to investigate the interaction effect on the condensed fraction. The condensed fraction $N_0/N$ as a function of the reduced temperature, with $\eta$ plays as a parameter is given. For all $\eta$, the condensed fraction decreases monotonically as the reduced temperature increases. Increasing $\eta$ leads to decreasing the condensed fraction $N_0/N$. For both $N/N_0$ and $T_C$, correction due to the finite size effect, which is responsible for the deviations from the ideal gas model, is about 2%. This effect is still visible for interacting gas, but it is strongly quenched in the presence of the interaction. Correction due to interaction effects is about 10%.

From Eq.(12) and (20) the specific heat per particle at temperature less than the transition temperature is given by

$$\left( \frac{C(t)}{Nk_B} \right)_{t<1} = 12 t^3 \frac{\zeta(4)}{\zeta(3)} + \frac{\bar{\omega}}{\Omega} t^2 \left\{ 9 \left( \frac{\zeta(3)}{N} \right)^{\frac{1}{3}} + 4\eta \left(1-t^3\right)^{\frac{3}{5}} (1 - \frac{7}{5} t^3) \right\} \qquad (22)$$

while at temperature greater than the transition temperature its given by

$$\left( \frac{C(t)}{Nk_B} \right)_{t>1} = 12 t^3 \frac{g_4(z)}{\zeta(3)} + \left( \frac{\bar{\omega}}{\Omega} \right) t^2 \left\{ 9 \frac{g_3(z)}{N^{\frac{1}{3}} \zeta(3)^{\frac{2}{3}}} + 4\eta \left(1-t^3\right)^{\frac{2}{5}} t^2 \frac{g_3(z)}{\zeta(3)} \right\} - \frac{\left[9 t^4 g_3(z) + 12 R t^3 g_2(z) + 4R^2 t^2 g_2^2(z)/\zeta(3)\right]}{t \, g_2(z) + R \, g_1(z)} \qquad (23)$$

where

$$R = \frac{\bar{\omega}}{\Omega} \left( \frac{\zeta(3)}{N} \right)^{\frac{1}{3}} \left\{ \frac{3}{2} + \frac{2}{3} \frac{\mu}{\hbar \Omega} \right\}$$

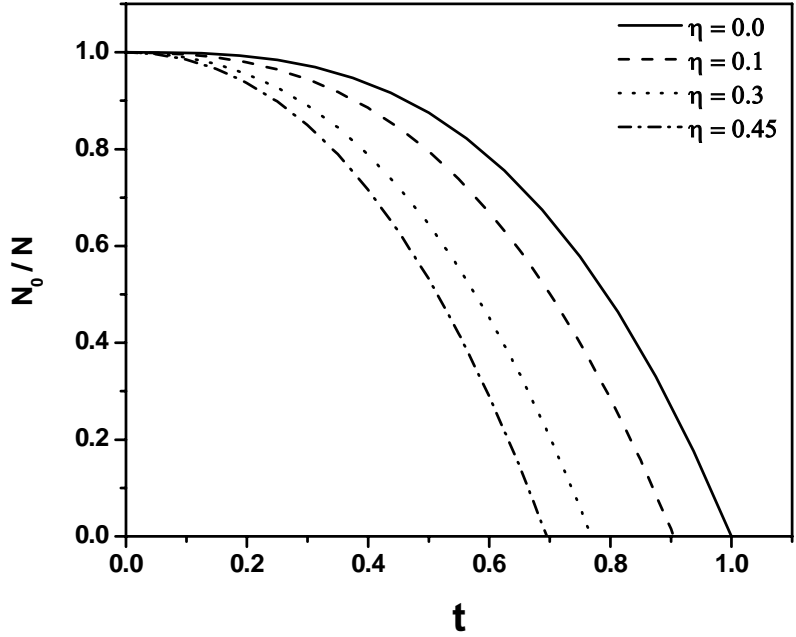

Figure 2: Condensed fraction $N_0/N$ as a function of the reduced temperature t, with η plays as a parameter. The results are calculated from Eq.(25).

Results calculated from Eq.(22) for the specific heat as a function of the reduced temperature t, with η used as a parameter is represented in figure (3). The η values are chosen to be relevant for the experimental values (= 0.1, 0.3, and 0.45 ). For all values of η, the specific heat increases monotonically with t until reaching a maximum value, and decreases very fast at temperature $T_0 > T > T_{max}$. From this figure one can see clearly that the specific heat has a maximum value for some certain temperature. This temperature can be identified as the temperature at which BEC occurs, we will refer to it by $T_{max}$. At temperature greater than $T_{max}$ the specific heat falls rapidly over a temperature range of about t ≈ 0.1. Remarkably, the specific heat becomes discontinuous at $T_0$. The magnitude of the jump is quite significant, in the thermodynamical limit it is given by

$$\frac{\Delta C^{(\infty)}}{Nk} = \frac{9t^4 g_3(z)}{t\, g_2(z)} - \frac{8}{5}\left(\frac{\bar{\omega}}{\Omega}\right)t^5\eta\,(1-t^3)^{-\frac{3}{5}} \qquad (24)$$

at t=1 the magnitude of the jump is given by, $(C(t)/NK_B)_{t=1} \approx 6.577$, i.e. full agreement with Grossmann et.al. result is recovered. Finally, the critical temperature $T_C$ can be obtained by setting $N_0 = 0$ in Eq.(21) [6],

$$T_c = T_0\left\{1 - 0.73\frac{\bar{\omega}}{\Omega}N^{-1/3} - 0.456\frac{\bar{\omega}}{\Omega}\eta(1-t^3)^{2/5}\right\} \qquad (25)$$

The first term on the right hand side of Eq.(25) gives exactly the usual results due to finite size effect. This effect is always negative and vanishes in the large N limit. The second term is the correction due to interaction, for $^{52}$Cr this term is negative. Griesmaier pointed that the experimental value for the critical temperature is given by $T_{exp}$ ~700nK.

**Effective width & Release energy**

The effective (square) width of a single particle state |n> of an interacting Bose gas at finite temperature is given by[9, 20, 21],

$$\langle r_n^2 \rangle = \frac{E_n}{\hbar\Omega}\sqrt{\frac{\hbar}{M\Omega}} = \frac{E_n}{\hbar\Omega}a_r^2 \qquad (26)$$

where $a_r$ is the characteristic length of the harmonic trap. Thus the square width of a trapped N atoms of a Bose gas is given by

$$<r^2> = \sum_{n=0}^{\infty} N_n \langle r_n^2 \rangle = \sum_{n=0}^{\infty} N_n \frac{E_n}{\hbar\Omega} a_r^2 = \frac{a_r^2}{\hbar\Omega} U$$

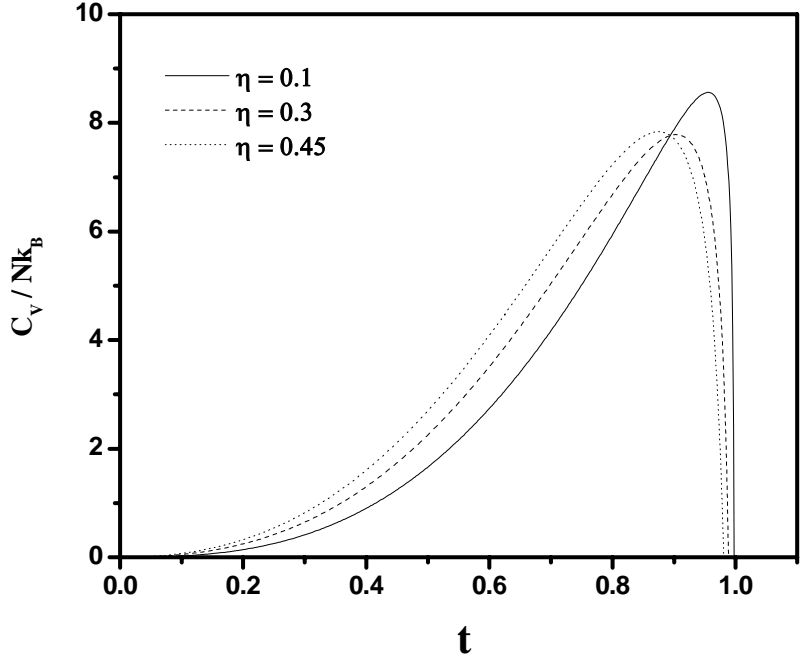

Figure 3: Specific heat per particle $C(t)/NK_B$ as a function of t with $\eta$ plays as a parameter. For $\eta = 0.1$; 0.3 and 0.45, the value of $t_{max}$ are 0.95, 0.90, and 0.87 respectively.

Where U is the total energy of the system. In terms of the grand potential we have

$$<r^2> \frac{a_r^2}{\hbar\Omega} K_B T^2 \frac{\partial q}{\partial T}|_z \qquad (27)$$

Substituting from Eq.(12) in (27) yields

$$\langle r^2 \rangle = a_r^2 \left\{ E_0 + 3a_r^2 \left(\frac{kT}{\hbar\Omega}\right)^4 g_4(z) + 2 a_r^2 \frac{\overline{\omega}}{\Omega} \left(\frac{kT}{\hbar\Omega}\right)^3 g_3(z) \left[\frac{3}{2} + \frac{2}{3}\frac{\mu}{\hbar\Omega}\right] \right\} \qquad (28)$$

Bracket in Eq.(28) takes a familiar form with the first term denoting the square width for the ground state (condensate), while the second term gives the excited states width (thermal component). The last term includes the correction due to finite size and interaction effects. Following Hassan et al. [9] and Zhang et al. [20], we have:

i) For temperature $T < T_0$, i.e. $t < 1$, the dimensionless square width is given by

$$\frac{\langle r^2 \rangle}{r_c^2} = t^4 + \left[ \frac{\zeta(3)}{\zeta(4)} \frac{\overline{\omega}}{\Omega} \left(\frac{\zeta(3)}{N}\right)^{\frac{1}{3}} + \frac{4}{9}\eta(1-t^3)^{2/5} \right] t^3 = t^3 \chi_1(t) \qquad (29)$$

With $r^2_c$ denotes to the square width of a Bose at temperature equal to the transition temperature $T_0$, $r^2_c = 3a^2_r \zeta(4) [N/\zeta(3)]^{4/3}$, and $\chi_1(t) = t + (\zeta(3)/\zeta(4)) (\omega/\Omega) [(\zeta(3)/N)^{1/3} + (4/9) \eta(1- t^3)^{2/5}]$. In the thermodynamic limit (N→∞ and $\eta$ → zero [23]) $\chi_1(t) \to t$.

ii) On the other hand for $T \geq T_0$, i.e. $t > 1$,

$$\langle r^2 \rangle \approx 3a_r^2 \left\{ N\frac{kT}{\hbar\Omega} \frac{g_4(z)}{\zeta(3)} + \frac{\overline{\omega}}{\Omega} \frac{N}{\zeta(3)} g_3(z) + \frac{2}{3} \frac{N}{\zeta(3)} g_3(z) \frac{\mu}{\hbar\Omega} \right\}$$

$$= \alpha r_c^2 t + \frac{\zeta(3)}{\zeta(4)} \frac{\overline{\omega}}{\Omega} \left[ \left(\frac{\zeta(3)}{N}\right)^{\frac{1}{3}} + \frac{4}{9}\eta(1-t^3)^{2/5} \right]$$

where $\alpha = (g_4(z)\zeta(3)) / (g_3(z)\zeta(4)) \approx 1$ has a very weak dependence on T. The dimensionless form is given by

$$\frac{\langle r^2 \rangle}{r_c^2} = \chi_1(t) \qquad (30)$$

We now come to our analysis of Eq.'s (29) & (30), and the fundamental difference between our results and the ideal Bose gas results given in Ref. [20]. An interesting feature is that the repulsive interaction causes the width of a Bose gas to increase at temperatures lower than the transition temperature ($T < T_0$), yet it has little effect on the width at temperatures higher than the transition temperature ($T > T_0$). The low temperature phenomenon is easy to understand in terms of a repulsive-interaction induced expansion of a Bose gas [12]. First, a condensate with repulsive interaction is larger in its size due to atom-atom repulsion; Second, the presence of a condensate pushes the thermal non-condensed cloud out, further increasing the size of a gas[13]. At high temperatures ($T > T_0$) the effect of repulsive interaction becomes negligible as the density of a Bose gas decreases dramatically with increasing temperatures. In figure 4, the temperature dependence effective width, $<r^2>^{1/2} / r_c$ is plotted. The solid, dashed, and dotted lines denote respectively the case of width with $\eta = 0$, 0.5, and 1.0, the number of particles is taken to be, $N = 1.3 \times 10^8$. There are a qualitative as will as a quantitative difference between the ideal Bose gas $\eta = 0.0$ and a system of interacting Bose gas. The effective size of a Bose gas $<r^2>$ is proportional to $T^4$ for $T < T_0$ and to T for $T \geq T_0$. This result consistent with earlier experimental reports that the area of absorption image of a Bose gas is proportional to its temperature in the absence of a condensate [24, 25].

Another important quantity is the release energy. In the experiments setup, the condensed Bose gas is produced at thermal equilibrium. When the trapping potential switches off suddenly, the cloud expands ballistically, and after a time long enough that the expansion velocity has reached a steady state value the kinetic energy of the expanding cloud is measured. This measured energy is known as the release energy. Since it is a pure kinetic energy, then it can be related to the effective size, $<r^2>$, through the relation[12, 21],

$$E_{rel} \approx E_{kin} \approx \frac{m}{2} \langle v^2 \rangle_{\tau \to \infty} \approx \frac{m}{2\tau^2} \langle r^2 \rangle_{\tau \to \infty} = \frac{m}{2\tau^2} r_c^2 \chi_1(t) \; t^3 \qquad (31)$$

where $\tau$ is the time of flight. The calculated results from Eq.(31) are illustrated in Figure 5. This figure reveals that the behavior of the release energy is similar to the behavior of the square width.

**Conclusion**

In this paper we have calculate measurable thermodynamical parameters for a trapped Boson gas. The analytical approach adopted in the present work would be effective to study the trapped interacting Bose gas. The main results emerging from our analysis are briefly summarized as follows:
i) interactions effect provide a significant quenching of the thermodynamical parameters with respect to the ideal gas model. Correction due to interaction is about 5 times the correction due to finite size effect. ii) A sudden drop for the effective width and expansion energy occur when temperature is lowered than the transition temperature. The sudden decreasing of the reduced width near the transition temperature has little dependence on the number of atoms. So, the effective width can serve as a good indication for the presence of BEC. In this site, Our results provide a solid theoretical foundation for the experiment. Finally, it would be straight-forward to generalize our findings for the DOS to a combined potential formed by an optical lattice potential and harmonic trap potential [26].

*Acknowledgments* : We are grateful to Professor Axel Griesmaier, physikalisches Institute, University of Stuttgart, Germany, for providing us with the experimental data.

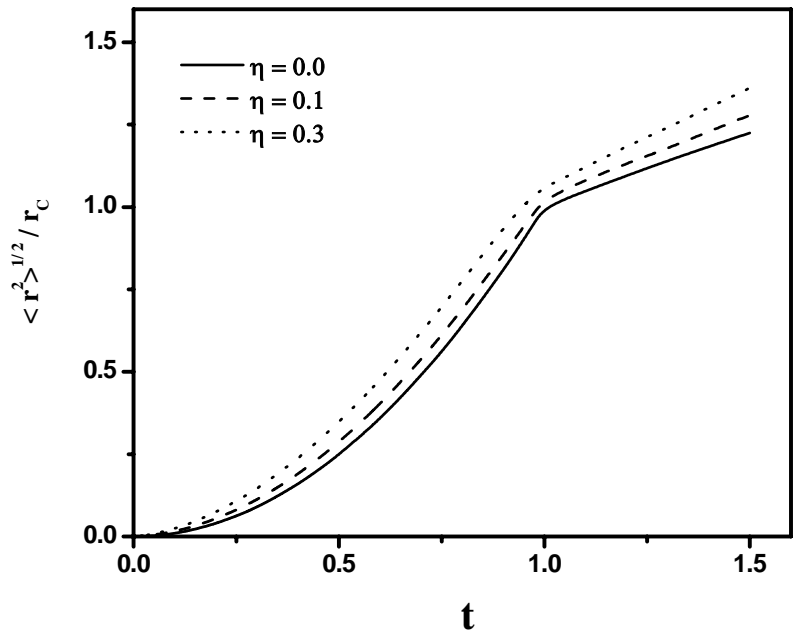

Figure 4: The temperature-dependence effective width for different interaction parameter.

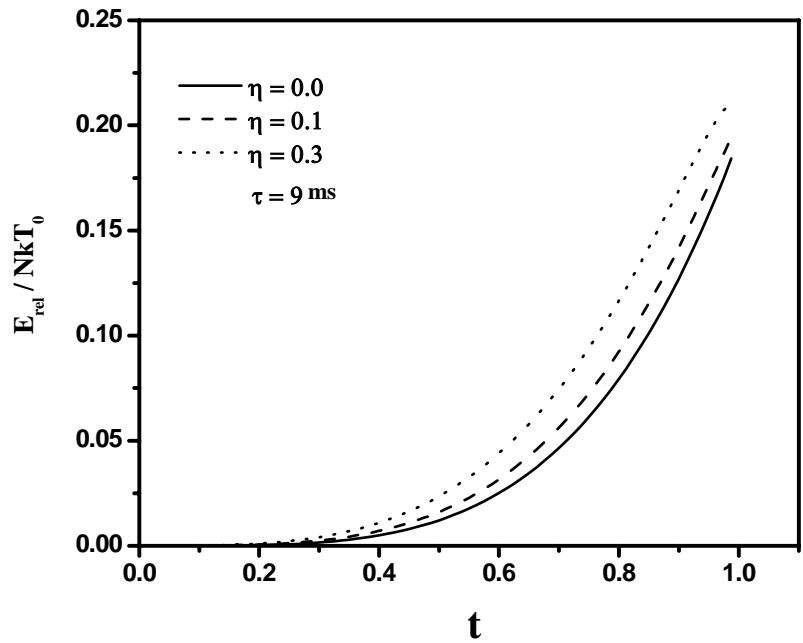

Figure 5: The release energy as a function of the reduced temperature.